\documentclass{revtex4-1}

\usepackage{graphicx}
\usepackage{bm}
\usepackage{float}

\begin{document}

\title{Spin Currents Induced by Nonuniform Rashba-Type Spin-Orbit Field}

\author{Kazuhiro Tsutsui}
\email[]{tsutsui@stat.phys.titech.ac.jp}
\affiliation{Department of Physics, Tokyo Institute of Technology, 2-12-1 Ookayama, Meguro-ku, Tokyo 152-8551, Japan}
\affiliation{Department of Physics, Tokyo Metropolitan University, Hachioji, Tokyo 192-0397, Japan}

\author{Akihito Takeuchi}
\affiliation{Department of Physics, Tokyo Metropolitan University, Hachioji, Tokyo 192-0397, Japan}

\author{Gen Tatara}
\affiliation{Department of Physics, Tokyo Metropolitan University, Hachioji, Tokyo 192-0397, Japan}

\author{Shuichi Murakami}
\affiliation{Department of Physics, Tokyo Institute of Technology, 2-12-1 Ookayama, Meguro-ku, Tokyo 152-8551, Japan}
\affiliation{PRESTO, Japan Science and Technology Agency (JST), Kawaguchi, Saitama 332-0012, Japan}

\begin{abstract}
We study the spin relaxation torque in nonmagnetic or ferromagnetic metals with nonuniform spin-orbit coupling within the Keldysh Green's function formalism. In non-magnet, the relaxation torque is shown to arise when the spin-orbit coupling is not uniform. In the absence of an external field, the spin current induced by the relaxation torque is proportional to the vector chirality of Rashba-type spin-orbit field (RSOF). In the presence of an external field, on the other hand, spin relaxation torque arises from  the coupling of the external field and vector chirality of RSOF. Our result indicates that spin-sink or source effects are controlled by designing RSOF in junctions.
\end{abstract}

\maketitle

\section{Introduction}

Generation of spin currents is of essential importance in spintronics.
So far, several methods have been known to generate spin currents, namely, spin injection from ferromagnetic metals, the spin Hall effect\cite{Hirsch99,Murakami03}, the spin pumping effect \cite{Silsbee79,Tserkovnyak02}, and the spin injection by the spin accumulation (spin chemical potential) \cite{Son87,Valet93}. 
It was recently demonstrated that thermal generation of spin currents is also possible (the spin Seebeck effect) \cite{Uchida08}.

In solids, the spin density ($ {\bm s}$) and spin current (${\bm j}_{\rm s}$)  satisfy the continuity equation 
with a source or sink term, ${\cal T}$;
\begin{equation}
\dot{\bm s}+\nabla\cdot{\bm j}_{\rm s}={\cal T}. 
\end{equation}
The right-hand side is the spin relaxation torque.
In metals, it arises mainly from the spin-orbit interaction.
In the case of uniform and dynamic magnetization, the relaxation torque reduces to the Gilbert damping torque \cite{KTS06,TE08}. 
In the context of the spin-transfer torque in inhomogeneous spin structures, the spin relaxation torque leads to a non-adiabatic torque ($\beta$ term) \cite{KTS06,TE08}.
In the case of electric spin injection from ferromagnetic metal to non-magnetic metal, the spin relaxation torque was shown to be proportional to $\nabla\cdot {\bm E}$ (${\bm E}$ is the applied electric field) and the spin injection effect is explained based on the behavior of  spin relaxation torque at the interface 
\cite{Nakabayashi10}.
The spin relaxation torque thus plays an essential role in spin current generation and dissipation.

To detect the spin current electrically, the inverse spin Hall effect is widely used by attaching metals with strong spin-orbit interaction, such as Pt  \cite{Saitoh06}. 
In the non-local spin injection experiment, 
Kimura \textit{et al.} discussed on a phenomenological ground that the spin current in Cu is absorbed by Pt due to the strong spin relaxation in Pt (spin sink effect), and that the absorbed spin current is converted into electric voltage by the inverse spin Hall effect \cite{Kimura07}. 
What is essential in the spin sink effect is a spatial inhomogeneity of the spin-orbit interaction, namely, the gradient of the spin-orbit interaction of the spin relaxation torque.
In this paper, we investigate this possibility by microscopic calculation.

\section{Model}

We consider the Hamiltonian with Rashba spin-orbit interaction coupled with an electromagnetic field. RSOF exists when the inversion symmetry is broken at the interface or surface \cite{Takeuchi10}. The total Hamiltonian ${\cal H}$ is composed of non-perturbative part ${\cal H}_0+{\cal H}_M+{\cal H}_{\mathrm{imp}}$ and perturbative part ${\cal H}^0_{\mathrm{SO}}+{\cal H}^A_{\mathrm{SO}}+{\cal H}_{\mathrm{em}}$. Each term is
\begin{eqnarray}
{\cal H}_0 &=& \sum_{\sigma=\pm 1} \int d\bm{r}\ c_\sigma^\dagger(\bm{r},t)\left(-\frac{\hbar^2}{2m}\nabla^2-\epsilon_F \right)c_\sigma(\bm{r},t), \\
{\cal H}_M &=& - \sum_{\sigma=\pm 1} \int d\bm{r}\ \sigma M c_\sigma^\dagger(\bm{r},t)c_\sigma(\bm{r},t), \\
{\cal H}_{\mathrm{imp}} &=&  \sum_{\sigma=\pm 1} \int d\bm{r}\ u(\bm{r}) c_\sigma^\dagger(\bm{r},t)c_\sigma(\bm{r},t), \\
{\cal H}^0_{\mathrm{SO}} &=& -i\hbar \sum_{\alpha,\beta=\pm1} \int d\bm{r}\ \bm{E}_{\mathrm{SO}}(\bm{r}) \cdot c_\alpha^\dagger(\bm{r},t) (\nabla\times \bm{\sigma}_{\alpha\beta}) c_\beta(\bm{r},t), \\
{\cal H}^A_{\mathrm{SO}} &=& e \sum_{\alpha,\beta=\pm1} \int d\bm{r}\ \bm{E}_{\mathrm{SO}}(\bm{r})\cdot c_\alpha^\dagger(\bm{r},t) (\bm{A}(\bm{r},t)\times \bm{\sigma}_{\alpha\beta}) c_\beta(\bm{r},t), \\
{\cal H}_{\mathrm{em}} &=& \frac{e\hbar}{2mi} \sum_{\sigma=\pm 1} \int d\bm{r}\ \bm{A}(\bm{r},t)\cdot c_\sigma^\dagger(\bm{r},t)\nabla c_\sigma(\bm{r},t),
\end{eqnarray}
where $c_\sigma(\bm{r},t)$ represents the annihilation operator of a conduction electron, $\bm{\sigma}_{\alpha\beta}$ represents Pauli matrices, $M$ is the magnetization, $\bm{E}_{\mathrm{SO}}(\bm{r})$ is RSOF and $\bm{A}(\bm{r},t)$ is a vector potential for the electromagnetic field.
$u(\bm{r})$ is the random impurity potential and averaging over the impurity position is carried out as $\langle u(\bm{r}) u(\bm{r}')\rangle_{\mathrm{imp}}=u_0^2 n_{\mathrm{imp}}\delta(\bm{r}-\bm{r}')$, where $u_0$ and $n_{\mathrm{imp}}$ are the strength of the impurity and the impurity concentration respectively.
The RSOF is assumed to arise from the gradient of a scalar potential and thus to satisfy $\nabla \times \bm{E}_{\mathrm{SO}}=\bm{0}$.
In the following, we calculate the spin torque in both systems, that is, non-magnets and ferromagnets, where we adapt $M=0$ for non-magnets and $M\neq 0$ for ferromagnets.

The spin density in the continuity equation is defined as $s^\alpha=\sum_{\beta \gamma} \langle c^\dagger_{\beta}\sigma^\alpha_{\beta \gamma}c_{\gamma} \rangle$, where $\beta$ and $\gamma$ represent the spin indices, and a spin current polarized in the $\alpha$ direction flowing in the $i$ direction is defined by
\begin{eqnarray}
(j_s^\alpha)_i &\equiv & \frac{1}{2m}\frac{\hbar}{i}\langle c^\dagger(\bm{r},t)\sigma^\alpha \partial_i c(\bm{r},t)\rangle -\frac{e}{m}A_i(\bm{r},t) \langle c^\dagger(\bm{r},t)\sigma^\alpha c(\bm{r},t)\rangle-\sum_j \epsilon_{j\alpha i} E^j_{\mathrm{SO}}(\bm{r}) \langle c^\dagger(\bm{r},t)c(\bm{r},t)\rangle.
\end{eqnarray}
The spin relaxation torque is derived by taking quantum and statistical averages of the Heisenberg's equation of motion for the spin operator \cite{TE08}. Its $\alpha$ component reads
\begin{equation}
{\cal T}^\alpha \equiv i\sum_{i,j,k}\sum_\gamma \epsilon_{ijk}\epsilon_{\alpha k\gamma} E^i_{\mathrm{SO}}(\bm{r})\langle c^\dagger(\bm{r},t)\sigma^\gamma \partial_j c(\bm{r},t) \rangle +\frac{2e}{\hbar}\sum_{i,j,k}\sum_\gamma \epsilon_{ijk}\epsilon_{\alpha k\gamma} E^i_{\mathrm{SO}}(\bm{r}) A_j(\bm{r},t) \langle c^\dagger(\bm{r},t)\sigma^\gamma c(\bm{r},t) \rangle,
\end{equation}
where $\epsilon_{ijk}$ stands for the Levi-Civita symbol.

\subsection{Equilibrium spin currents}

By calculating the spin relaxation torque perturbatively within the Keldysh Green's function formulation, we access the issue of spin source and sink. The contribution of spin relaxation torque which is i-th order in the RSOF and j-th order in the vector potential is denoted as ${\cal T}_{(i,j)}^\alpha$. First, we deal with the case without the vector potential, i.e., ${\cal T}_{(i,0)}^\alpha$. The first order contribution with respect to the RSOF, ${\cal T}_{(1,0)}^\alpha$, vanishes. The second order contribution is calculated as
\begin{eqnarray}
{\cal T}_{(2,0)}^\alpha &=& i \frac{\hbar}{2}\sum_{i,j,k}\sum_\gamma \sum_{l,m,n} \epsilon_{ijk}\epsilon_{\alpha k\gamma}\epsilon_{lmn} \int \frac{d\omega}{2\pi} \nonumber \\
 & & \sum_{\bm{k}}\sum_{\bm{p}}\sum_{\bm{p}_1} e^{i(\bm{p}+\bm{p}_1)\cdot \bm{r}} (2k-p_1)_j(2k+p_1)_m E^i_{\mathrm{SO}}(\bm{p})E^l_{\mathrm{SO}}(\bm{p}_1) \mathrm{tr}\left[ \sigma^\gamma \hat{g}_{\bm{k},\omega} \sigma^n \hat{g}_{\bm{k}-\bm{p}_1,\omega} \right]^<,
\end{eqnarray}
where $\hat{g}_{\bm{k},\omega}$ denotes Green's function of free conduction electrons in the wave-number space. Trace in the spin space is represented by $\mathrm{tr}$ and $[\ ]^<$ denotes the lesser component.

\subsubsection{Non-magnets}

We first consider non-magnets, where $\hat{g}_{\bm{k},\omega}=g_{\bm{k},\omega}\hat{1}$ ($\hat{1}$ is the identity matrix of spin space).
By using the rotational symmetry, the angular average of the wave vectors is calculated as $\langle k_ik_jk_lk_m \rangle=\frac{1}{15}k^4(\delta_{ij}\delta_{lm}+\delta_{il}\delta_{jm}+\delta_{im}\delta_{jl})$.
We see that the leading contribution contains the second order derivative of the RSOF.
The torque is calculated as
\begin{equation}
{\cal T}_{(2,0)}^\alpha=i \frac{4\hbar}{15}\left( \frac{\hbar^2}{m}\right)^2 \int \frac{d\omega}{2\pi}\sum_{\bm{p},\bm{p}_1} e^{i(\bm{p}+\bm{p}_1)\cdot \bm{r}} \sum_{\bm{k}} k^4 \left[ (g^A_{\bm{k},\omega})^4-(g^R_{\bm{k},\omega})^4 \right] \left( \bm{E}_{\mathrm{SO}}\times \bm{p}_1^2\bm{E}_{\mathrm{SO}}(\bm{p}_1) \right)^\alpha,
\end{equation}
where $g^A_{\bm{k},\omega}$ represents the advanced Green's function and $g^R_{\bm{k},\omega}$ is the retarded one.
The result is
\begin{equation}
{\cal T}_{(2,0)}^\alpha=\frac{2\pi \hbar}{15} \frac{\nu}{\epsilon_F} \left( \bm{E}_{\mathrm{SO}}\times \nabla^2 \bm{E}_{\mathrm{SO}} \right)^\alpha, \label{T20}
\end{equation}
where $\nu$ represents the density of states.
As we see, the spin relaxation torque of eq. (\ref{T20}) is zero when the RSOF is spatially uniform. The relaxation torque of eq. (\ref{T20}) generates, when the system is static, the spin current given by
\begin{equation}
(\bm{j}^\alpha_s)_i=\frac{2\pi \hbar}{15} \frac{\nu}{\epsilon_F} \left[ \bm{E}_{\mathrm{SO}}\times (\partial_i \bm{E}_{\mathrm{SO}}) \right]^\alpha.
\end{equation}
This implies that vector chirality of the RSOF, $\bm{E}_{\mathrm{SO}}\times (\partial_i \bm{E}_{\mathrm{SO}})$, produces a spin current. 

\begin{figure}
 \begin{center} 
 \includegraphics[width=80mm]{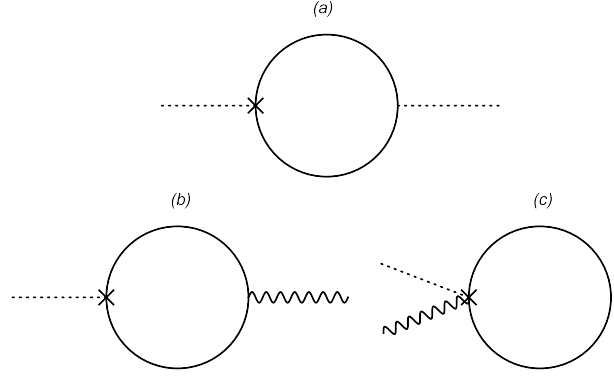} 
 \end{center} 
 \caption{Diagrammatic representation of the contribution of the spin relaxation torque. Contribution (a) is the second order contribution in the RSOF. Contribution (b) and (c) are the first order in the RSOF (dotted lines) and linear in the external field (wavy lines).} 
 \label{so2so1em1} 
 \end{figure}

\subsubsection{Ferromagnets}

In the case of spin-polarized conduction electrons, i.e., in ferromagnets, the spin relaxation torque is affected by spin polarization.
Nevertheless, the contributions of the zeroth and first order derivative of the RSOF vanish in the same way as in unpolarized case, resulting in
\begin{equation}
{\cal T}_{(2,0)}^\alpha=\frac{\pi \hbar}{15} \left( \frac{\nu_+}{\epsilon^+_F}+\frac{\nu_-}{\epsilon^-_F} \right) \left( \bm{E}_{\mathrm{SO}}\times \nabla^2 \bm{E}_{\mathrm{SO}} \right)^\alpha. \label{T20}
\end{equation}
Here, $\nu_\sigma\ (\sigma=\pm)$ is the density of states for spin-split bands and $\epsilon^\sigma_F\equiv \epsilon_F-\sigma M$.
Under equilibrium condition, whether the spin of conduction electrons is polarized or not is not so important for the qualitative behavior of the spin relaxation torque.

\subsection{Non-equilibrium spin currents}

We study in this subsection the contribution linear in the external field.

\subsubsection{Ferromagnets}

First, let us discuss the spin-polarized conduction electrons. In this case, the leading contribution of spin relaxation is the first order with respect to RSOF as shown in Figs. \ref{so2so1em1} (b) and  \ref{so2so1em1} (c). The spin relaxation torque then reads
\begin{eqnarray}
{\cal T}_{(1,1)}^\alpha &=& i \frac{e\hbar}{2m}\sum_{i,j,k}\sum_\gamma\sum_l \epsilon_{ijk}\epsilon_{\alpha k\gamma} \int \frac{d\omega}{2\pi}\int \frac{d\Omega}{2\pi} e^{-i\Omega t} \nonumber \\
 & & \times \sum_{\bm{k}}\sum_{\bm{p}}\sum_{\bm{q}} e^{i(\bm{p}+\bm{q})\cdot \bm{r}} (2k-p)_j(2k+q)_l\ A_l(\bm{q},\Omega) E^i_{\mathrm{SO}}(\bm{p}) \mathrm{tr} \left[ \sigma^\gamma \hat{g}_{\bm{k},\omega} \hat{g}_{\bm{k}-\bm{q},\omega-\Omega} \right]^<.
\end{eqnarray}
Using $\langle k_ik_j \rangle=\frac{1}{3}k^2\delta_{ij}$ as a result of rotation symmetry, the leading contribution of the torque is
\begin{equation}
{\cal T}_{(1,1)}^\alpha=i\frac{2e\hbar}{3m}\int \frac{d\Omega}{2\pi} \ \Omega e^{-i\Omega t} \sum_{\bm{p},\bm{q}}e^{i(\bm{p}+\bm{q})\cdot \bm{r}} \sum_{\bm{k},\sigma} \sigma k^2 g^R_{\bm{k},\sigma} g^A_{\bm{k},\sigma} \Big[ (\bm{E}_{\mathrm{SO}}(\bm{p})\times \bm{A}(\bm{q},\Omega)) \times \bm{e}_z \Big]^\alpha.
\end{equation}
Here, $\bm{e}_z$ denotes the unit vector of $z$ direction.
We obtain
\begin{equation}
{\cal T}_{(1,1)}^\alpha=\frac{8\pi}{3}\frac{e}{\hbar^2} \left( \tau_+ \epsilon^+_F \nu_+ -\tau_- \epsilon^-_F \nu_- \right) \Big[ (\bm{E}_{\mathrm{SO}}\times \bm{E}_{\mathrm{em}}) \times \bm{e}_z \Big]^\alpha,
\end{equation}
where $\tau_\sigma (\sigma=\pm)$ is the relaxation time for spin-split bands.
As we see, the spin relaxation torque as a linear response of an external field arises even in the uniform Rashba field case. This spin relaxation may be the dominant mechanism of spin sinks in the junction of a ferromagnet and a non-magnet where $E_{\mathrm{em}}$ is applied along the ferromagnet and the interface Rashba exists perpendicular to $E_{\mathrm{em}}$.
We note that contributions containing derivatives such as in Eq. (12) arise also in ferromagnets as so small corrections to ${\cal T}_{(1,1)}^\alpha$.

\subsubsection{Non-magnets}

\begin{figure}
 \begin{center} 
 \includegraphics[width=80mm]{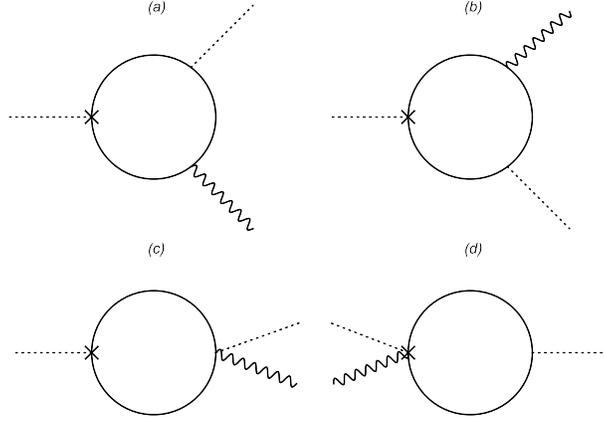} 
 \end{center} 
 \caption{Diagrammatic representation of the contribution of the second order in the RSOF and linear in the external field.} 
 \label{so2em1} 
 \end{figure}

If we assume that $A(\bm{r},t)=A(t)$ for simplicity, the dominant torque in the spin-unpolarized case is diagrammatically given by (a) and (b) in Fig. \ref{so2em1} and is described as
\begin{eqnarray}
{\cal T}_{(2,1)}^\alpha &=& i \frac{e\hbar^2}{4m}\sum_{i,j,k}\sum_\gamma \sum_{l,m,n}\sum_o \epsilon_{ijk}\epsilon_{\alpha k\gamma}\epsilon_{lmn} \int \frac{d\omega}{2\pi}\int \frac{d\Omega}{2\pi} e^{-i\Omega t} \sum_{\bm{k}}\sum_{\bm{p}}\sum_{\bm{p}_1,\bm{q}} e^{i(\bm{p}+\bm{p}_1+\bm{q})\cdot \bm{r}} \nonumber \\
 &\times& \left[ (2k-p_1-q)_j (2k+p_1)_m (2k-2p_1+q)_o\ E^i_{\mathrm{SO}}(\bm{p})E^l_{\mathrm{SO}}(\bm{p}_1)A_o(\bm{q},\Omega) \mathrm{tr} \left[ \sigma^\gamma \hat{g}_{\bm{k},\omega} \sigma^n \hat{g}_{\bm{k}-\bm{p}_1,\omega}\hat{g}_{\bm{k}-\bm{p}_1-\bm{q},\omega-\Omega} \right]^< \right. \nonumber \\
 & & \left.+(2k-p_1-q)_j (2k-2q+p_1)_m (2k+q)_o\ E^i_{\mathrm{SO}}(\bm{p})E^l_{\mathrm{SO}}(\bm{p}_1)A_o(\bm{q},\Omega) tr\left[ \sigma^\gamma g_{\bm{k},\omega} g_{\bm{k}-\bm{q},\omega-\Omega}\sigma^n g_{\bm{k}-\bm{q}-\bm{p}_1,\omega-\Omega} \right]^< \right].
\end{eqnarray}
By noting $\nabla \times \bm{E}_{\mathrm{SO}}=\bm{0}$ and $\nabla \times \bm{E}_{\mathrm{em}}=\bm{0}$, the leading contribution of the torque turns out to be the forth order in wave number and the first order in the derivative of RSOF, resulting in
\begin{eqnarray}
{\cal T}_{(2,1)}^\alpha &=& -\frac{4e}{15}\left( \frac{\hbar^2}{m}\right)^2 \int \frac{d\Omega}{2\pi} \ \Omega e^{-i\Omega t} \sum_{\bm{p},\bm{p}_1} e^{i(\bm{p}+\bm{p}_1)\cdot \bm{r}} \ \Im \left(\sum_{\bm{k}} k^4 (g^R_{\bm{k}})^4\right) \nonumber \\
 & & \times \Big[ \left( \bm{E}_{\mathrm{SO}}(\bm{p})\times (\bm{A}(\Omega)\cdot \bm{p}_1)\bm{E}_{\mathrm{SO}}(\bm{p}_1) \right)^\alpha+ \bm{A}(\Omega)\cdot \left( \bm{E}_{\mathrm{SO}}(\bm{p})\times \partial_\alpha \bm{E}_{\mathrm{SO}}(\bm{p}_1)\right) \Big].
\end{eqnarray}
Here, $g^R_{\bm{k}}\equiv g^R_{\bm{k},\omega=0}$.
We obtain
\begin{equation}
{\cal T}_{(2,1)}^\alpha=\frac{\pi e}{15}\frac{\nu}{\epsilon_F} \Big[ \left( \bm{E}_{\mathrm{SO}}\times (\bm{E}_{\mathrm{em}}\cdot \nabla) \bm{E}_{\mathrm{SO}} \right)^\alpha +\bm{E}_{\mathrm{em}}\cdot (\bm{E}_{\mathrm{SO}}\times \partial_\alpha \bm{E}_{\mathrm{SO}}) \Big]. \label{lo}
\end{equation}
The structure of the first term of the right hand side of eq.(\ref{lo}) is the same as that of so-called $\beta$ term in the current-induced torque, $-\beta (\bm{S}\times (\bm{j}\cdot \nabla)\bm{S})$, if we replace the localized spin $\bm{S}$ by $\bm{E}_{\mathrm{SO}}$. The second term represents coupling of the electromagnetic field and the vector chirality of RSOF.

Let us look in detail into the spin relaxation torque of eq. (\ref{lo}) for a junction of two nonmagnetic metals shown in Fig. 3 (a). As seen from eq. (\ref{lo}),  the RSOF at the interface needs to form a vector chirality (${\bm E}_{\rm SO}\times \partial {\bm E}_{\rm SO}$)  to induce finite spin relaxation torque.
Such configuration is expected to arise generally when the junction is of finite area.
Since the Rashba effect is due to the structure-inversion-symmetry breaking of a quantum well and the RSOF arises from the gradient of a scalar potential, the RSOF will behave near the edges as shown in Fig. 3 (b).
Let us consider a square junction area in the $xy$ plane with an applied electric field (or current) along the $y$ direction.
Then both the first and the second terms of the right-hand side of eq. (\ref{lo}) have only the component with $\alpha=x$.
The sign of the relaxation torque is negative along the $x$ direction and positive along $y$ direction, resulting in the torque density distribution of Figs. 3 (e) and 3 (f) with the spin sink along the $x$ axis and spin source along the $y$ axis.
In the static case, the spin current generated by the present torque density is polarized in the $x$ direction and satisfies $\nabla\cdot {\bm j}_{\rm s}^x={\cal T}^x$.
In the analogy with the Gauss's law of the classical electromagnetism, the generated-spin-current distribution is as depicted in Figs. 3 (c), 3 (d) and 3 (g).
Since the induced torque comes from the vector chirality of the RSOF, the spin-current distribution generated by the spin relaxation torque is the same for the opposite sides of the system, which is different from that induced by the spin Hall effect.
Experimental observation of the local spin current would be interesting.

As we have discussed, the spin continuity equation is useful to understand the spin current generation and absorption based on the spin relaxation torque distribution. One should note, however, that the result of the spin current determined this way has an ambiguity given by a rotation of some vector, $\nabla\times {\bm C}$ (${\bm C}$ is any vector field). It would be important to investigate other equations for the spin current defining its rotation. (In the case of static classical electromagnetism, the electric field  is constrained by the rotation free condition, $\nabla\times {\bm E}=0$.)

\begin{figure}
 \begin{center} 
 \includegraphics[width=130mm]{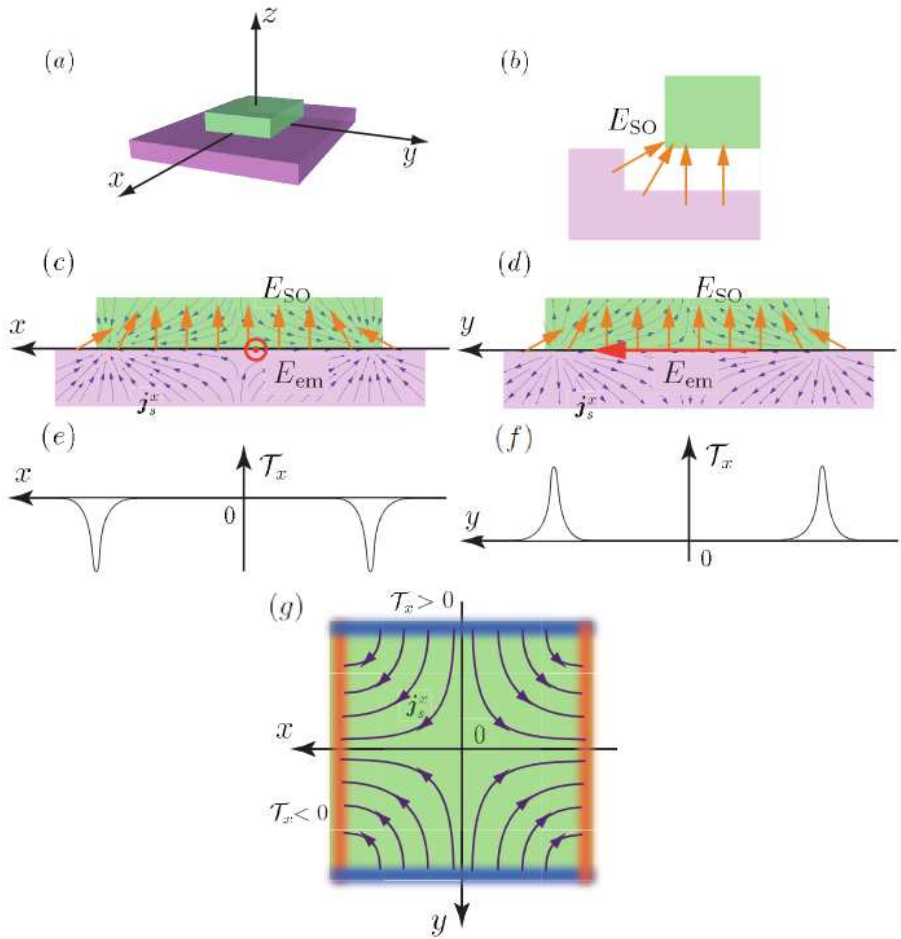} 
 \end{center} 
 \caption{(Color online) Schematic figures of spin currents induced by the spin relaxation torque of eq. (\ref{lo}). RSOF is induced in the interface between the upper non-magnet and the lower one. Yellow arrows denote the RSOF ($\bm{E}_{\mathrm{SO}}$), a red arrow denotes an external field ($\bm{E}_{\mathrm{em}}$) and purple arrows denote spin currents ($\bm{j}^x_s$). Orange (blue) region indicates plus (minus) values of $x$-component spin relaxation torque. (a): A square junction in the $xy$ plane. (b): The spatial distribution of the RSOF near the edge. (c) and (d): Configuration along the $x$ axis and $y$ axis respectively. (e) and (f): The spatial distribution of $x$-component spin relaxation torque. (g): The spin-currents distribution induced by the spin relaxation torque.} 
 \label{SS} 
 \end{figure}

\section{Summary}

In summary, we presented a microscopic theory of spin relaxation torque in nonmagnetic and ferromagnetic metals with nonuniform spin-orbit coupling within the Keldysh Green's function formalism. We found that the spin relaxation torque depends quantitatively on whether an external electric field is applied or not and on the spin polarization of the carrier. In the case of non-magnets, there exist equilibrium spin currents induced by vector chirality of RSOF in the absence of an external field. In the presence of an external field, on the other hand, spin relaxation torque arises from the coupling of the electric field and vector chirality of RSOF. In ferromagnets, in contrast, uniform RSOF results in the spin relaxation torque. Our result indicates that control of RSOF will make possible the spin-currents control without magnetism or magnetic field.

\begin{acknowledgments}
KT is grateful for J. Yoshino, K. Taguchi and N. Nakabayashi. This work is supported by a Grant-in-Aid for Scientific Research in Priority Areas (Grant No. 1948027) from The Ministry of Education, Culture, Sports, Science and Technology (MEXT) and a Grant-in-Aid for Scientific Research (B) (Grant No. 22340104) from Japan Society for the Promotion of Science (JSPS).
\end{acknowledgments}

\end{document}